\documentclass[a4paper,pra,amsmath,amssymb,superscriptaddress,preprintnumbers,onecolumn,12pt]{revtex4}

\usepackage{graphicx,color}

\begin{document}

\title{Observation of an Efimov-like resonance in ultracold atom-dimer scattering}

\author{S. Knoop$^*$}
\affiliation{Institut f\"ur Experimentalphysik und Zentrum f\"ur
Quantenphysik, Universit\"at
  Innsbruck, 6020 Innsbruck, Austria}

\author{F. Ferlaino}
\affiliation{Institut f\"ur Experimentalphysik und Zentrum f\"ur
Quantenphysik, Universit\"at
  Innsbruck, 6020 Innsbruck, Austria}

\author{M. Mark}
\affiliation{Institut f\"ur Experimentalphysik und Zentrum f\"ur
Quantenphysik, Universit\"at
  Innsbruck, 6020 Innsbruck, Austria}

\author{M. Berninger}
\affiliation{Institut f\"ur Experimentalphysik und Zentrum f\"ur
Quantenphysik, Universit\"at
  Innsbruck, 6020 Innsbruck, Austria}

\author{H. Sch\"{o}bel}
\affiliation{Institut f\"ur Experimentalphysik und Zentrum f\"ur
Quantenphysik, Universit\"at
  Innsbruck, 6020 Innsbruck, Austria}

\author{H.-C. N\"{a}gerl}
\affiliation{Institut f\"ur Experimentalphysik und Zentrum f\"ur
Quantenphysik, Universit\"at
  Innsbruck, 6020 Innsbruck, Austria}

\author{R. Grimm}
\affiliation{Institut f\"ur Experimentalphysik und Zentrum f\"ur
Quantenphysik, Universit\"at
  Innsbruck, 6020 Innsbruck, Austria}
\affiliation{Institut f\"ur Quantenoptik und Quanteninformation,
\"Osterreichische Akademie der Wissenschaften, 6020 Innsbruck,
Austria\\$^{*}$e-mail: steven.knoop@uibk.ac.at}
\date{\today}

\maketitle

{\bf The field of few-body physics has originally been motivated by
understanding nuclear matter. New model systems to experimentally
explore few-body quantum systems can now be realized in ultracold
gases with tunable interactions
\cite{Kraemer2006efe,Braaten2006uif}. Albeit the vastly different
energy regimes of ultracold and nuclear matter (peV as compared to
MeV), few-body phenomena are universal for near-resonant two-body
interactions \cite{Braaten2006uif}. Efimov states represent a
paradigm for universal three-body states \cite{Efimov1970ela}, and
evidence for their existence has been obtained in measurements of
three-body recombination in an ultracold gas of caesium atoms
\cite{Kraemer2006efe}. Interacting samples of halo dimers
\cite{Ferlaino2008cbt} can provide further information on universal
few-body phenomena. Here we study interactions in an optically
trapped mixture of such halo dimers with atoms, realized in a
caesium gas at nanokelvin temperatures. We observe an atom-dimer
scattering resonance, which we interpret as being due to a trimer
state hitting the atom-dimer threshold. We discuss the close
relation of this observation to Efimov's scenario
\cite{Efimov1970ela}, and in particular to atom-dimer Efimov
resonances \cite{Efimov1979lep,Nielsen2002eri,Braaten2007rdr}.}

Ultracold quantum gases offer an unprecedented level of control and
are versatile systems to investigate interacting quantum systems.
Their unique property is that the two-body interaction, as described
by the $s$-wave scattering length $a$, can be magnetically tuned
through Feshbach resonances
\cite{Tiesinga1993tar,Inouye1998oof,Courteille1998ooa}. When $|a|$
is tuned to values much larger than the range of the two-body
potential, one enters the universal regime \cite{Braaten2006uif},
where details of the short range interaction become irrelevant
because of the long-range nature of the wavefunction. For ground
state alkali atoms the range of the interaction potential is
determined by the van der Waals interaction and is given by $r_{\rm
vdW} = \frac{1}{2}(2\mu C_6/\hbar^2)^{1/4}$, where $C_6$ is the van
der Waals dispersion coefficient and $\mu$ is the reduced mass. A
manifestation of universality in two-body physics is the existence
of a \emph{quantum halo} dimer state at large positive $a$, for
which the binding energy is given by the universal expression
$E_{\rm b}=\hbar^2/(2\mu a^2)$ \cite{Jensen2004sar}. In the field of
ultracold gases such halo dimers, with $a\gg r_{\rm vdW}$ and
$E_{\rm b}\ll E_{\rm vdW}= \hbar^2/(2\mu r_{\rm vdW}^2)$, can be
experimentally realized \cite{Kohler2006poc,Ferlaino2008cbt}, with
the unique possibility to tune the properties of the halo dimers
across the universal and non-universal regimes. The concept of
universality extends beyond two-body physics to few-body phenomena
\cite{Braaten2006uif}. So far pure atomic systems have been used to
experimentally study universal three-body physics. By introducing
halo dimers a new class of phenomena becomes accessible, including
universal atom-dimer and dimer-dimer scattering.

\begin{figure}
\includegraphics[width=8.5cm]{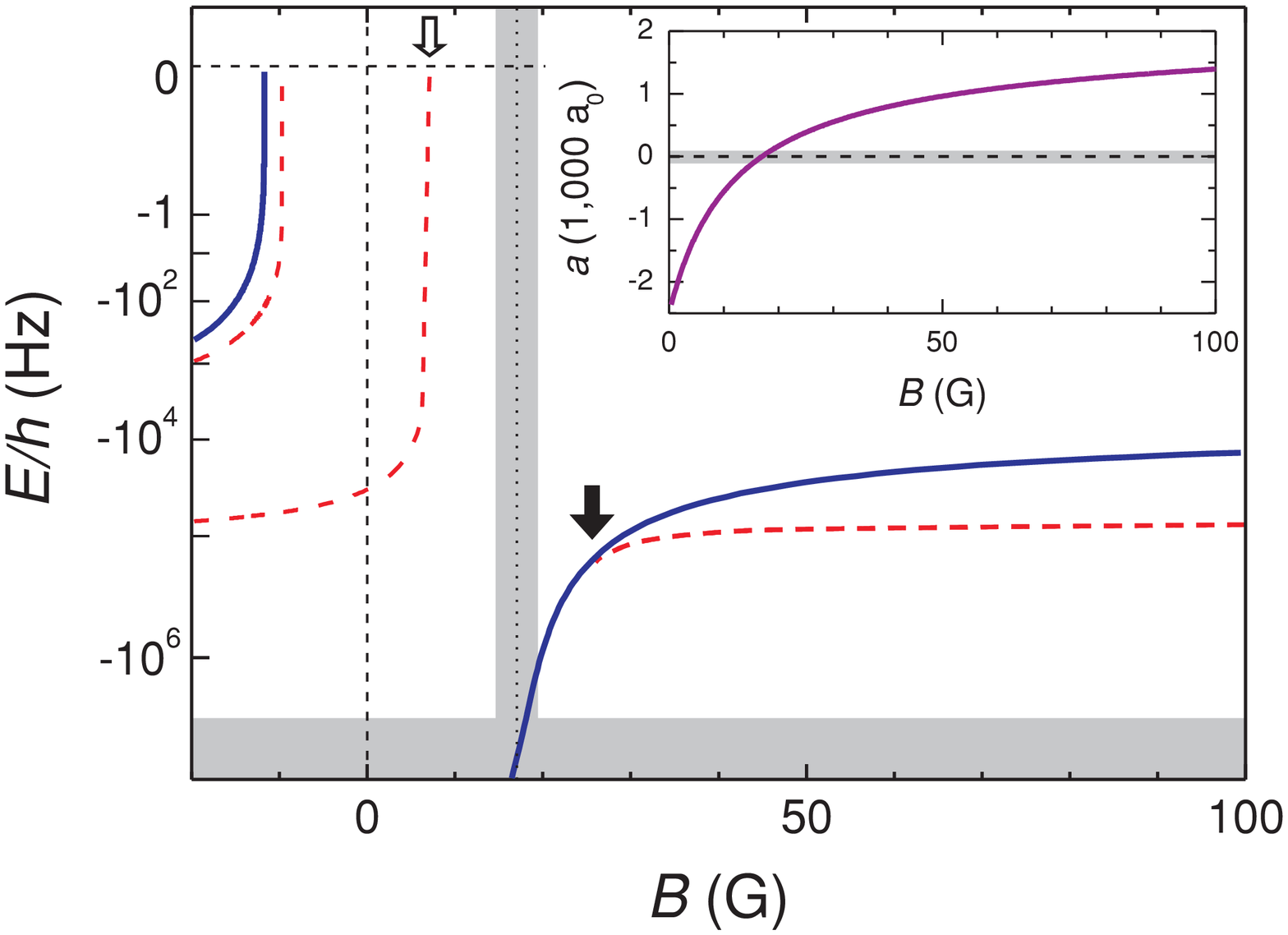}
\caption{{\bf Three-body spectrum of caesium.} The energies $E$ of
the atom-dimer thresholds (blue solid curves) are shown as function
of the magnetic field $B$. The red dashed lines illustrate
Efimov-like trimer states, for which the energy dependence is not
precisely known. The giant three-body loss resonance found at 7.5~G
\cite{Kraemer2006efe} has pinpointed the intersection of an Efimov
state with the three-atom threshold (open arrow). The intersection
of an Efimov state with an atom-dimer threshold (filled arrow) leads
to a resonance in atom-dimer relaxation. Zero energy corresponds to
three atoms in the lowest spin state, labeled by the total spin
quantum number $F=3$ and its projection $m_F=3$. The inset shows the
scattering length $a$ as a function of the magnetic field $B$. The
grey areas represent the non-universal regions, where $|a|<r_{\rm
vdW}=100\,a_0$ or $E_b>E_{\rm
vdW}=h\times2.7$~MHz.}\label{fig:threebodypicture}
\end{figure}

Here we focus on three-body systems consisting of three identical
bosons. In particular, we consider the situation in which two atoms
are bound, forming a halo dimer, and scatter with a third, free
atom. The atom-dimer system under investigation consists of
$^{133}$Cs atoms in their lowest spin state, labeled by the total
spin quantum number $F=3$ and its projection $m_F=3$. The caesium
atoms represent an excellent system to study few-body physics with
bosons at large scattering lengths because of their unique
scattering properties \cite{Chin2004poc}. The scattering length $a$
at low magnetic fields is shown in the inset of
Fig.~\ref{fig:threebodypicture}. Its unusual magnetic-field
dependence is explained by a broad Feshbach resonance at a magnetic
field of $-12$~G \cite{Weber2003tbr,Kraemer2006efe} along with an
extraordinarily large background scattering length and allows to
tune $a$ from large negative to positive values with a zero crossing
at 17~G. Note that a negative magnetic field corresponds to a sign
reversal of the projection quantum number $m_F$.

The three-body energy spectrum is shown in
Fig.~\ref{fig:threebodypicture}, illustrating the energies of trimer
states (red dashed curves) and atom-dimer thresholds (blue solid
curves). Here zero energy corresponds to three free atoms with zero
kinetic energy. The energy of an atom-dimer threshold thus simply
corresponds to the binding energy of a dimer. The energy
dependencies of these thresholds are well known, because of the
precise knowledge the caesium two-body spectrum
\cite{Chin2004poc,Mark2007sou}. The dimer state that corresponds to
the atom-dimer threshold at positive magnetic fields carries halo
character as $a\gg r_{\rm vdW}$ is fulfilled in a wide
magnetic-field range; for caesium $r_{\rm vdW}=100\,a_0$, where
$a_0$ is Bohr's radius, and $E_{\rm vdW}=h\times2.7$~MHz. Below 20~G
the dimer state bends downwards into the non-universal region
because of an avoided crossing with the dimer state that causes the
Feshbach resonance at $-12$~G.

The trimer states, as illustrated in
Fig.~\ref{fig:threebodypicture}, are located in the regime where
$|a|$ exceeds $r_{\rm vdW}$, with binding energies well below
$E_{\rm vdW}$. We therefore refer to them as Efimov states
\cite{Efimov1970ela}, although sometimes more strict definitions are
used \cite{Lee2007ete}. An Efimov trimer intersects the three-atom
threshold, at which three free atoms couple resonantly to a trimer.
Similarly, an Efimov trimer couples to a halo dimer and a free atom
at the atom-dimer threshold. The energy spectrum of trimer states is
not precisely known, but their appearance at the thresholds can give
clear signatures of their locations. The observation of a giant
three-body recombination loss resonance in an ultracold atomic
caesium sample at 7.5~G, corresponding to $a=-850\,a_0$, has
pinpointed the location at which one of the Efimov states hits the
three-atom threshold \cite{Kraemer2006efe}; see open arrow. The next
Efimov resonance in three-body recombination loss is predicted at
negative magnetic fields, in principle accessible with atoms in the
$F=3$, $m_F=-3$ state. Unfortunately, in practice its observation
will be obscured by fast two-body losses \cite{Gueryodelin1998ibe}.
Several studies have suggested the intersection of a trimer state
with the atom-dimer threshold for positive magnetic fields below
50~G \cite{Koehler2007privat,Esry2007privat,Massignan2008esn}; see
filled arrow. The appearance of an Efimov trimer at the atom-dimer
threshold is predicted to manifest itself in a resonance in
atom-dimer relaxation \cite{Nielsen2002eri,Braaten2007rdr}.

Atom-dimer relaxation is energetically possible because of the
presence of deeply bound dimer states, providing many inelastic
channels for collisions between atoms and weakly bound dimers. This
process leads to loss of both the atom and the dimer from the trap,
as the acquired kinetic energy is in general much larger than the
trap depth. The particle loss is described by the rate equation
$\dot{n}_{\rm D}=\dot{n}_{\rm A}=-\beta n_{\rm D} n_{\rm A}$, where
$n_{\rm D}$($n_{\rm A}$) is the molecular (atomic) density and
$\beta$ denotes the loss rate coefficient for atom-dimer relaxation.
In the non-universal regime, relaxation loss in ultracold atom-dimer
samples has been studied in various systems
\cite{Wynar2000mia,Mukaiyama2004dad,Staanum2006eio,Zahzam2006amc,Syassen2006cdo,Hudson2008ico}
and was found to be essentially independent of the magnetic field.
In the universal regime, suppression of loss has been observed in
systems involving fermions \cite{Inguscio2008ufg,Zirbel2008cso}.

\begin{figure}
\includegraphics[width=8.5cm]{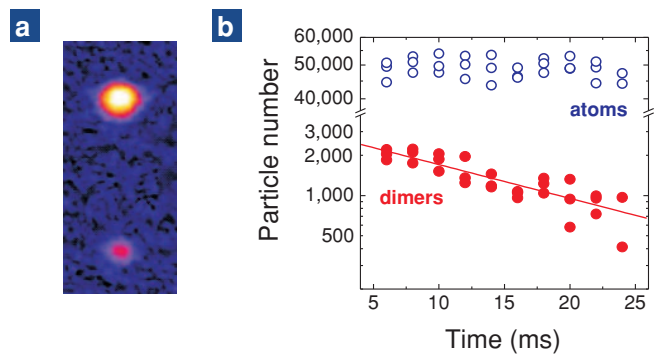}
\caption{{\bf Measuring the atom-dimer relaxation loss rate.} {\bf
a,} Absorption image of the atom-dimer mixture after release from
the trap and Stern-Gerlach separation. {\bf b,} Time evolution of
the number of atoms and dimers at 35~G. Here the loss of dimers can
be fitted with an exponential decay curve with a $1/e$ lifetime
proportional to $\beta^{-1}$, as the atom number greatly exceeds the
dimer number and loss due to the dimer-dimer relaxation can be
neglected (see Methods).}\label{fig:decaycurves}
\end{figure}

Starting point of our measurements is an ultracold atom-dimer
mixture prepared in a crossed-beam optical dipole trap in the
temperature range of 30-250~nK  (see Methods section). The dimers
are created by Feshbach association
\cite{Herbig2003poa,Kohler2006poc} at a 200-mG wide Feshbach
resonance at 48~G (not shown in Fig.~\ref{fig:threebodypicture}) and
then transferred into the halo dimer state
\cite{Mark2007sou,Ferlaino2008cbt}. For our lowest temperatures we
obtain a mixture of about $3\times10^{4}$ atoms and $4\times10^{3}$
dimers. After preparation of the mixture we ramp to a certain
magnetic field and wait for a variable storage time. Then we switch
off the trap and let the sample expand before ramping back over the
48-G resonance to dissociate the molecules, after which standard
absorption imaging is performed. During the expansion a magnetic
field gradient is applied to spatially separate the atomic and
molecular cloud (Stern-Gerlach separation) \cite{Herbig2003poa}. In
this way we simultaneously monitor the number of atoms and dimers,
see Fig.~\ref{fig:decaycurves}a. A typical loss measurement is shown
in Fig.~\ref{fig:decaycurves}b. We observe a fast loss of dimers on
the timescale of a few tens of a millisecond. In order to obtain
$\beta$ we have set up a dimer loss model based on the
above-mentioned rate equation (see Methods section). Because the
number of atoms greatly exceeds the number of dimers, a simple
analytic expression can be derived, which is fitted to the data. The
loss of dimers due to dimer-dimer relaxation is small and is taken
into account; the corresponding loss rate for this process was
measured independently using a pure dimer sample
\cite{Ferlaino2008cbt}.

\begin{figure}
\includegraphics[width=8.5cm]{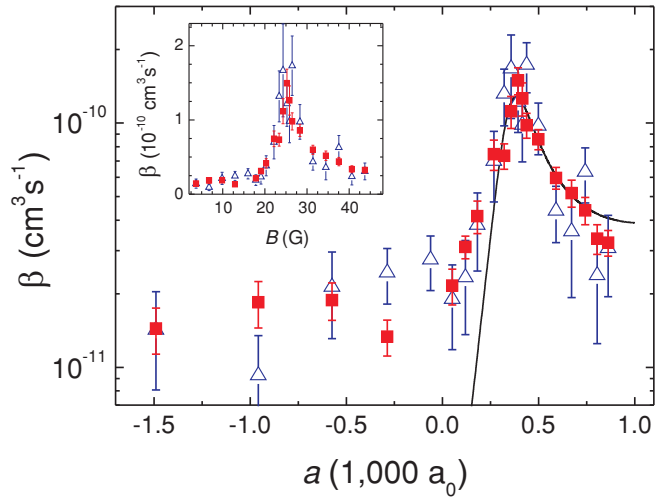}
\caption{{\bf Loss resonance in atom-dimer relaxation.} The loss
rate coefficient $\beta$ for atom-dimer relaxation is shown as a
function of the scattering length $a$ (main figure) and the magnetic
field $B$ (inset); measurements are taken at temperatures of
40(10)~nK (blue open triangles) and 170(20)~nK (red closed squares).
The solid curve is a fit of an analytic model from effective field
theory (see text) to the data for $a>r_{\rm
vdW}=100\,a_0$.}\label{fig:lossrate}
\end{figure}

The resulting $\beta$ is shown in Fig.~\ref{fig:lossrate} as a
function of the two-body scattering length $a$, using the magnetic
field dependence of the scattering length (inset of
Fig.~\ref{fig:threebodypicture}); the inset shows the same data as a
function of the magnetic field. For $a<0$ ($B<17$~G) we observe an
essentially constant $\beta$ of about
$1.5\times10^{-11}$~cm$^3$s$^{-1}$, similar to other observations in
the non-universal regime. A constant $\beta$ can be understood by
the fact that in this region the dimers are non-universal and
therefore their properties are not directly connected to the
scattering length \cite{Kohler2006poc,Ferlaino2008cbt}. For $a>0$ we
observe a resonance at about $a=400\,a_0$ ($B=25$\,G), where $\beta$
increases by one order of magnitude to
$1.5\times10^{-10}$~cm$^3$s$^{-1}$. Two data sets at different
temperatures are shown, namely at 40(10)~nK (blue open triangles)
and 170(20)~nK (red closed squares). Both data sets give the same
behaviour regarding resonance position and loss rate, showing that
the measurements are in the threshold regime in which the loss rate
is independent of the collision energy and not unitarity limited. We
interpret the observed resonance as the expected manifestation of a
trimer state hitting the atom-dimer threshold.

Even though the position of the resonance is not far in the
universal regime, we compare our findings with the prediction from
effective field theory in the universal limit \cite{Braaten2007rdr}.
We consider the analytic expression $\beta=C_{\rm AD}(a)\hbar a/m$
with $C_{\rm
AD}(a)=D\left[\sinh(2\eta_*)/\left(\sin^2\left[s_0\ln(a/a_*)\right]+\sinh^2\eta_*\right)\right]$,
where $s_0=1.00624$. The logarithmically periodic dependence of
$C_{\rm AD}(a)$ on $a$ is characteristic for Efimov physics. Here
$a_*$ describes the position at which the Efimov trimer hits the
atom-dimer threshold, defined up to a factor
$e^{\pi/s_0}\approx22.7$. The dimensionless quantity $\eta_*$ is
related to the lifetime of Efimov states near threshold, and thus
characterizes the width of the resonance. We have fitted the
analytic expression to the 170~nK data for $a>r_{\rm vdW}=100 a_0$,
with $a_*$, $\eta_*$ and $D$ as free parameters. The result is shown
as the solid curve in Fig.~\ref{fig:lossrate}. We find that the
analytic expression describes the data well, in particular the
asymmetric shape of the resonance is well represented by the
universal prediction. For the resonance position and the width
parameter we obtain $a_*=367(13)\,a_0$ and $\eta_*=0.30(4)$,
respectively. Both parameters are free in the framework of
Ref.~\cite{Braaten2007rdr}. For $D$ we find $2.0(2)$, for which
Ref.~\cite{Braaten2007rdr} predicts a fixed value of 20.3.

According to the universal theory, observations on atom-dimer
relaxation are connected to those in three-body recombination
\cite{Braaten2006uif}. The three-body recombination loss rate can be
expressed as $L_3=3C_\pm(a)\hbar a^4/m$, where $C_+(a)$ ($a>0$) and
$C_-(a)$ ($a<0$) are again logarithmically periodic functions of $a$
(see \cite{Kraemer2006efe} and references herein). From a fit to the
three-body recombination data a maximum in $C_+(a)$ was found at
$a_+=1060(70)\,a_0$ \cite{Kraemer2006efe}. Universal theory predicts
$a_*\approx1.1a_+$ from which an atom-dimer resonance would be
expected around $1,200\,a_0$. We find the resonance at a smaller
value of $a$, implying that here an obvious universal connection is
absent. The universal theory also connects Efimov-related features
for $a<0$ and $a>0$. The three-body recombination resonance at
$a_-=-850(20)\,a_0$ would correspond to an atom-dimer resonance
around $900\,a_0$, also significantly shifted from our present
finding. This may be related to the fact that the transition from
$a<0$ to $a>0$ takes place via a zero crossing and not via a pole as
in the Efimov scenario \cite{Kraemer2006efe,DIncao2007iit}. A
universal connection was suggested by the three-body recombination
data \cite{Kraemer2006efe}, but in view of the new observations this
might have been fortuitously \cite{DIncao2007iit}.

The observation of a resonance shows that atom-dimer relaxation
measurements can provide information on weakly bound trimer states
in a complementary way to three-body atomic recombination. Our
observations in three-body systems involving caesium follow
qualitatively the Efimov scenario, with a three-body recombination
resonance at $a<0$ and an atom-dimer relaxation resonance at $a>0$.
However, the relation between the different features is
quantitatively not well described by the universal theory. A simple
explanation might be that $a$ is not large enough for a
straight-forward application of predictions valid in the universal
limit. This raises the question whether it is possible to extend
universal theory by incorporating non-universal corrections
\cite{Hammer2007erc} or whether a fully non-universal theory would
be required.

Further experiments around Feshbach resonances at higher magnetic
fields, in particular around a broad one at 800\,G
\cite{Lee2007ete}, can provide deeper insight into the universal
aspects of three-body physics. First of all, in the universal theory
the locations of three-body recombination and atom-dimer relaxation
features only depend on the scattering length. Under this assumption
our observations in the low magnetic field range will reappear at
higher magnetic fields at the same values of the scattering length.
Secondly, at the Feshbach resonances the regions with $a<0$ and
$a>0$ are connected via a pole in $a$, and the universal relation
between these two regions can be tested. Finally, much larger values
of scattering length will be accessible, allowing to probe Efimov
physics much deeper in the universal regime.

\section*{Methods}\label{methods}

\subsection*{Preparation and detection}

Our ultracold atom-dimer mixture is trapped in a crossed-beam
optical dipole trap generated by two 1,064-nm laser beams with
waists of about 250~$\mu$m and 36~$\mu$m \cite{Ferlaino2008cbt}.
Since atoms and dimers in general have different magnetic moments
the application of a levitation field is not appropriate and a
sufficiently high optical gradient in the vertical direction to hold
the atoms and dimers against gravity is required. However, to obtain
very low temperatures and not too high densities a tight trap is not
advantageous. Here we use an adjustable elliptic trap potential with
weak horizontal confinement and tight confinement in the vertical
direction. The ellipticity is introduced by a rapid spatial
oscillation of the 36-$\mu$m waist beam in the horizontal plane with
the use of an acousto-optic modulator, creating a time-averaged
optical potential. The final temperature of the atomic and molecular
sample can be set by varying the ellipticity and the laser power of
the laser beam in the final trap configuration. For the lowest
temperature samples, the final time-averaged elliptic potential is
characterized by trap frequencies of 10~Hz and 20~Hz in the
horizontal plane, and 80~Hz in the vertical direction.

\subsection*{Dimer loss model}

We measure the atom-dimer relaxation loss rate $\beta$ by recording
the time evolution of the dimer number $N_{\rm D}$ and atom number
$N_{\rm A}$. In a harmonic trap the atomic and molecular samples can
be described by Gaussian density distributions, where the width
depends on the trap frequencies, the temperature and the mass.
Because the polarizability of the halo dimers is twice that of the
atoms, the trap frequencies of the atoms and the dimers are the
same. We find that the atomic and molecular samples have the same
temperature \cite{Ferlaino2008cbt}. The time-evolution of $N_D$ can
then be described by the following rate equation:
\begin{equation}\label{moleculeequation}
\dot{N}_{\rm D}=-\frac{8}{\sqrt{27}} \beta \bar{n}_{\rm A} N_{\rm
{\rm D}}-\alpha \bar{n}_{\rm D} N_{\rm D},
\end{equation}
with $\bar{n}_{\rm A}=[m\bar{\omega}^2/(4\pi k_{\rm B}
T)]^{3/2}N_{\rm A}$ and $\bar{n}_{\rm D}=[m\bar{\omega}^2/(2\pi
k_{\rm B} T)]^{3/2}N_{\rm D}$ the mean atomic and molecular density,
respectively, $m$ the atomic mass, $\bar{\omega}$ the geometric mean
of the trap frequencies and $T$ the temperature. Here loss of dimers
due to dimer-dimer relaxation is also taken into account via the
dimer-dimer relaxation loss rate coefficient $\alpha$. Because of
the unequal mass, the density distributions of the atomic and
molecular samples are not the same. As a result, an effective atomic
density experienced by the molecular cloud has to be considered,
which is taken into account by the factor $\frac{8}{\sqrt{27}}$ in
front of the atom-dimer loss term \cite{Staanum2006eio}.

Our experiments are carried out in the regime in which $N_{\rm
A}$$\gg$$N_{\rm D}$ and loss of atoms as a result of atom-dimer
relaxation is negligible. Three-body recombination leads to atom
loss on a much longer timescale compared to the molecular lifetime
\cite{Weber2003tbr}. Therefore $N_{\rm A}$ can be taken as a
constant and equation~(\ref{moleculeequation}) has the following
solution:
\begin{equation}\label{solutionmoleculeequation}
N_{\rm D}(t)=\frac{bN_{\rm A} N_{\rm D,0}}{(bN_{\rm A}+aN_{\rm
D,0})e^{bN_{\rm A}t}-aN_{\rm D,0}},
\end{equation}
where $N_{\rm D,0}\equiv N_{\rm D}(t=0)$,
$b\equiv\frac{8}{\sqrt{27}}\beta[m\bar{\omega}^2/(4\pi k_{\rm B}
T)]^{3/2}$ and $a\equiv\alpha[m\bar{\omega}^2/(2\pi k_{\rm B}
T)]^{3/2}$. If $\beta N_{\rm A}$$\gg$$\alpha N_{\rm D}$, i.e.
dimer-dimer relaxation loss is negligible compared to atom-dimer
relaxation loss, equation~(\ref{solutionmoleculeequation})
simplifies to
\begin{equation}\label{solutionmoleculeequation2}
N_{\rm D}(t)=N_{\rm D,0}e^{-bN_{\rm A}t},
\end{equation}
and $N_{\rm D}$ shows an exponential decay with a $1/e$ lifetime of
$(bN_{\rm A})^{-1}$. In our experiments dimer-dimer relaxation loss
can be neglected for $B>20$\,G and
equation~(\ref{solutionmoleculeequation2}) is fitted to the data.
For $B<20$\,G $\beta$ is much smaller than $\alpha$
\cite{Ferlaino2008cbt} and the application of
equation~(\ref{solutionmoleculeequation}) is required, taking
$\alpha$ from independent loss measurements of a pure dimer sample
\cite{Ferlaino2008cbt}.

For each measurement of $\beta$ the trap frequencies and the
temperature are determined by sloshing mode and time-of-flight
measurements, respectively. The error bars on $\beta$ contain the
uncertainties of these trap frequencies and temperature
measurements.

\section*{Acknowledgements}

We thank T. K\"{o}hler, B. Esry and P. Massignan for many fruitful
discussions. We acknowledge support by the Austrian Science Fund
(FWF) within SFB 15 (project part 16). S.~K.\ is supported within
the Marie Curie Intra-European Program of the European Commission.
F.~F.\ is supported within the Lise Meitner program of the FWF.

\section*{Competing financial interests}
The authors declare no competing financial interests.

\clearpage

\end{document}